\documentclass[aps,prl,twocolumn,superscriptaddress]{revtex4}

\usepackage{graphicx}
\usepackage[german,english]{babel}
\begin{document}
\title{First-principles studies of water adsorption on graphene: The role of the substrate}

\author{Tim O. Wehling}
\email{twehling@physnet.uni-hamburg.de}
\affiliation{I. Institut f{\"u}r Theoretische Physik, Universit{\"a}t Hamburg, Jungiusstra{\ss}e 9, D-20355 Hamburg, Germany}
\author{Mikhail I. Katsnelson}
\affiliation{Institute for Molecules and Materials, Radboud
University of Nijmegen, Heijendaalseweg 135, 6525 AJ Nijmegen, The
Netherlands}
\author{Alexander I. Lichtenstein}
\affiliation{I. Institut f{\"u}r Theoretische Physik, Universit{\"a}t Hamburg, Jungiusstra{\ss}e 9, D-20355 Hamburg, Germany}


\date{\today}

\begin{abstract}
We investigate the electronic properties of graphene upon water
adsorption and study the influence of the SiO$_2$ substrate in
this context using density functional calculations. Perfect suspended graphene is
rather insensitive to H$_2$O adsorbates, as doping requires highly
oriented H$_2$O clusters. For graphene on a defective SiO$_2$ substrate, we
find a strongly different behavior: H$_2$O adsorbates can shift
the substrate's impurity bands and change their hybridization with
the graphene bands. In this way, H$_2$O can lead to
doping of graphene for much lower adsorbate
concentrations than for free hanged graphene. The
effect depends strongly on the microscopic substrate properties.
\end{abstract}

\maketitle

Graphene, i.e. a monolayer of graphite, is the first truly two
dimensional (2D) material
\cite{Novoselov_science2004,K.S.Novoselov07262005} and a promising
candidate for silicon replacement in semiconductor industry
\cite{morozov08} or gas sensing applications
\cite{schedin-gassensors,WehlingNL08}. Since graphene's discovery
the water adsorbates have been discussed as impurities leading to
doping \cite{Novoselov_science2004,schedin-gassensors}, while
changing the electron mobility in graphene only surprisingly
little. Up to now, the microscopic mechanism of this doping
without significant changes in electron mobility has remained
unclear. Density functional theory calculations on single water molecule
adsorbates on perfect free standing graphene \cite{leenaerts} were
in line with previous studies on carbon nanotubes (CNT)
\cite{Zhao_CNT_H2O_02} and found H$_2$O physisorption but no
H$_2$O induced impurity states close to the Fermi level.
Therefore, the doping effects found experimentally
\cite{Novoselov_science2004,schedin-gassensors} are very likely due
to more complicated mechanisms than interactions of graphene with
single water molecules. The experiments dealing with the effect of
water on graphene were carried out using graphene on top of
SiO$_2$ substrates. In addition, for finite concentrations of H$_2$O on
graphene H$_2$O clusters might form. Despite its importance for understanding the doping experiments \cite{Novoselov_science2004,schedin-gassensors} and for possible graphene based applications, a microscopic theory on the role of the substrate and {H$_2$O}-clusters in the {H$_2$O}-graphene-interplay has been lacking, so far.

\begin{figure}[ht]
\centering
\includegraphics{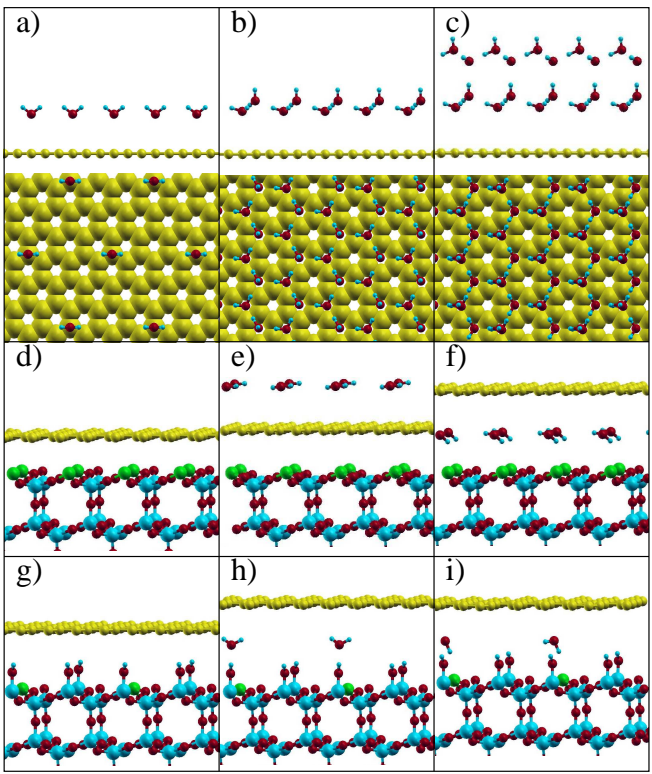}
\caption{\label{fig:geom} (Color online) Model systems for water interacting with graphene. a-c: Free standing graphene with single water adsorbates (a), a bilayer (b) and a tetralayer (c) of ice Ih. Carbon atoms yellow, oxygen red and hydrogen small blue balls. d-i: graphene on top of SiO$_2$ with every second (d-f) or eighth (g-i) surface Si atom forming a Q$_3^0$ defect. Water adsorbates are considered on top of graphene (e), and between graphene and the substrate (f), (h), and (i). Fully coordinated Si atoms are depicted as big blue balls, Si atoms at Q$_3^0$ defects in green.}
\end{figure}

In this letter we study the substrate and cluster formation
effects in the water-graphene-system by means of density
functional theory (DFT). We show that both, highly oriented water
clusters as well as water adsorbates in combination with a
defective SiO$_2$ substrate can lead to doping of graphene. To this end, we consider model systems (see Fig. \ref{fig:geom}) for water and ice in different concentrations of free standing graphene as well as for water interacting with defective SiO$_2$ substrates. By
analyzing the involved dipole moments and comparison to
electrostatic force microscopy \cite{bachtold}, we show that the
SiO$_2$ substrate is crucial for obtaining doping by H$_2$O adsorbates on
graphene.

In our DFT calculations, the electronic and structural properties
of the graphene-substrate-adsorbate systems are obtained using
generalized gradient approximation (GGA)\cite{Perdew:PW91, PBE} to
the exchange correlation potential. For solving the resulting
Kohn-Sham equation we used the Vienna Ab Initio Simulation Package
(VASP) \cite{Kresse:PP_VASP} with the projector augmented wave
(PAW) \cite{Bloechl:PAW1994,Kresse:PAW_VASP} basis sets
implemented, there. The graphene-substrate-adsorbate systems are
modeled using supercells containing up to 83 substrate atoms (Si,
O and H), 12 adsorbate atoms and 32 C atoms.

\begin{figure*}[t]
\centering
 \includegraphics{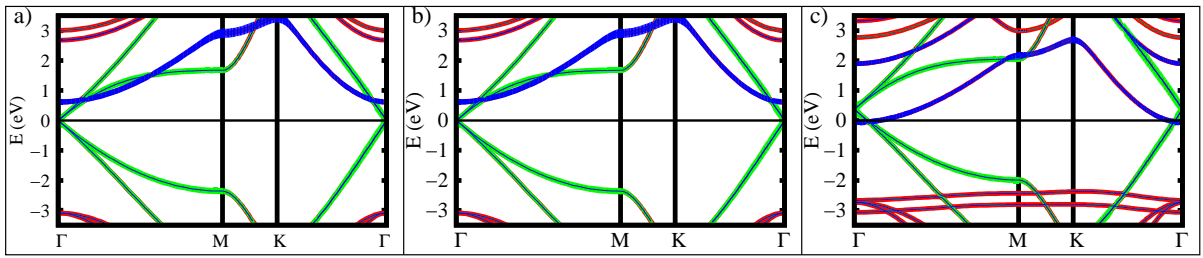}
\caption{\label{fig:Pure_H2O_ol_perf}(Color online) Bandstructures of supercells with fully relaxed single molecules (a), a bilayer (b) and a tetralayer (c) of water on graphene, corresponding to Fig. \ref{fig:geom} a-c), respectively, are shown. The graphene $\pi$ bands are marked in green, the nearly free electron bands in blue. Due to the {H$_2$O} dipole moments, graphene's nearly free electron band is shifted with respect to its $\pi$ bands.}
\end{figure*}

Firstly, water adsorption on free standing graphene with different
water concentrations is considered. To model a single H$_2$O
molecule on graphene, $3\times 3$ graphene supercells were used.
Full relaxation of H$_2$O with oxygen or hydrogen nearest to the
graphene yielded adsorbed configurations with binding energies of
$40$\,meV and $36$\,meV and molecule sheet distances of $3.50$\,{\AA}
and $3.25$\,{\AA}, respectively. These values are in the same order
as those obtained by Leenaerts et. al. \cite{leenaerts} indicating
physisorption of single water molecules on graphene. Analyzing the
density of states/band structures for these two adsorption
geometries we find qualitative agreement with Ref.
\cite{leenaerts}: None of these configurations exhibits energy
levels due to the adsorbate near the Dirac point, as shown in Fig.
\ref{fig:Pure_H2O_ol_perf} a) for the configuration with oxygen
closest to graphene. The HOMO of {H$_2$O} is more than $2.4$\,eV
below the Fermi energy and its LUMO more than $3$\,eV above it.
The absence of any additional impurity level close to the Dirac
point shows that single water molecules on perfect free standing
graphene sheets do not cause any doping.

The supercell applied here, corresponds to an adsorbate
concentration of $n=2$\,nm$^{-2}$, which is well inside the range
of concentrations (1-10 nm$^{-2}$) found experimentally in Ref.
\cite{bachtold}. The lateral dimension $a=4.5$\,{\AA} of the
hexagonal ice Ih (0001)-surface unit cell corresponds to a
concentration of $5.7$\, H$_2$O\,nm$^{-2}$ per layer. Thus,
increasing the {H$_2$O} concentration significantly above the
$n=2$\,nm$^{-2}$ from above leads to water clusters or ice like structures, rather than
isolated molecules.

To gain insight into doping of graphene by water clusters and ice
overlayers we studied fully relaxed bi-layers and four layers of ice
Ih adsorbed on graphene. These overlayer structures have been
proposed as basis of ice growth on various hexagonal metal
surfaces \cite{Lankau:2004,Kaxiras07} and can be modeled as
$(\sqrt{3}\times\sqrt{3})$\,R$30^\circ$ overlayer on the simple
graphene unit cell. The lattice mismatch in this configuration is
$0.23$\,{\AA} --- on the same order as found for water overlayers on
Ni(111) \cite{Lankau:2004} --- and therefore a reasonable
starting point for studying ice on graphene.

The supercell bandstructures (Fig. \ref{fig:Pure_H2O_ol_perf} b)
and c)) show that the electric field by proton-ordered ice on top
of graphene changes the energy of graphene's nearly free electron
bands. In contrast to pristine graphene, where these bands start
at 3eV above the Dirac point or in the case of single {H$_2$O}
adsorbates on graphene (Fig. \ref{fig:Pure_H2O_ol_perf} a)), their
bottom is at 0.6eV above and 0.1eV below the Dirac point for a bi-
and a tetralayer of ice Ih on top of graphene. This shift is due to electrostatic fields induced by the {H$_2$O} dipole moments and results in hole doping for the tetralayer of ice on graphene.

The water adlayers cause a change in contact potential $\delta
\phi$, which can be estimated using the H$_2$O dipole moment
$p=6.2\times 10^{-30}$\,Cm and the relaxed structures to be
$\delta \phi=1.4$V and 5.4V for a water bi- and tetralayer,
respectively. While only the latter structure causes doping, the
corresponding change in contact potential exceeds the experimental
value, $p_{\rm exp}=1.3$\,V, from Ref. \cite{bachtold} by more than a factor of
4. However, water strongly diluted in N$_2$ has been found to
cause hole doping in graphene on SiO$_2$\cite{schedin-gassensors}.
Given these two experiments, doping due to multiple fully oriented
ice overlayers as in Fig. \ref{fig:Pure_H2O_ol_perf} is likely not
the most important interaction mechanism for water and graphene.

We now turn to studying the effect of the SiO$_2$ substrate in the
water-graphene interplay. The experiments with graphene on top of
SiO$_2$ used substrates, which were created by plasma oxidation of
Si. \cite{schedin-gassensors,bachtold} The SiO$_2$ surface created
in this way is amorphous and its electronic, structural and
chemical properties are challenging to model from first
principles. To obtain, qualitative insight to the most important
physical mechanisms it is, however, a reasonable starting point to
consider crystalline SiO$_2$ in the $\beta$-cristobalite form as a
substrate \cite{cristobalite}.

The (111) surface of this modification can be used to create the
most likely defects on SiO$_2$ amorphous surfaces: These are so
called $Q_3^0$ and $Q_4^1$ defects \cite{wilson:9180} having one
under coordinated silicon and oxygen atom, respectively.
Furthermore, the unit cell of this surface is nearly commensurate
with the graphene lattice: The lattice constant $a_{{\rm
SiO}_2}=7.13$\,{\AA} \cite{cristobalite_wright} results in a surface
unit cell $a_{{\rm SiO}_2}/\sqrt{2}=5.04$\,{\AA}, which is  $4\%$
larger than twice the lattice constant $2a_0=4.93$\,{\AA} of
graphene. As the SiO$_4$ tetrahedra in SiO$_2$ are known to adjust
to external strain easily, we model graphene on SiO$_2$ by
$2\times 2$ or $4\times 4$ graphene supercells with lateral
dimension $2a_0$ or $4a_0$. As substrate we put $4\%$ laterally
strained and hydrogen passivated $\beta$-cristobalite with 6 Si
atoms per surface unit cell in vertical direction. We then created
the defects by removing H passivation atoms, added the H$_2$O
adsorbates and relaxed until all forces were less than
$0.08$\,eV$\cdot${\AA}$^{-1}$. In this way, we consider passivated and
defective SiO$_2$ surfaces --- the latter containing either
undercoordinated silicon or oxygen atoms --- as substrate for
graphene. The effect of water exposure is simulated by putting
water molecules on top of graphene as well as between graphene and
the substrate.

Graphene on top of fully passivated SiO$_2$ means two inert systems in
contact with each other. Consequentely, there are no bands in addition to
graphene's Dirac bands at the Fermi level and no doping occurs.
(The band structure is not shown here, for brevity.) This changes
strongly for graphene on defective SiO$_2$. As a model system, we
study $Q_3^0$ defects in $\beta$-cristobalite (111)-surfaces.
Depending on the supercell size, $2\times 2$ and $4\times 4$, in
these periodic structures every second and eighth surface Si atom
is under coordinated, respectively, and forms a $Q_3^0$ defect. (See Fig. \ref{fig:geom} (d-i).)

\begin{figure*}[t]
\centering
 \includegraphics{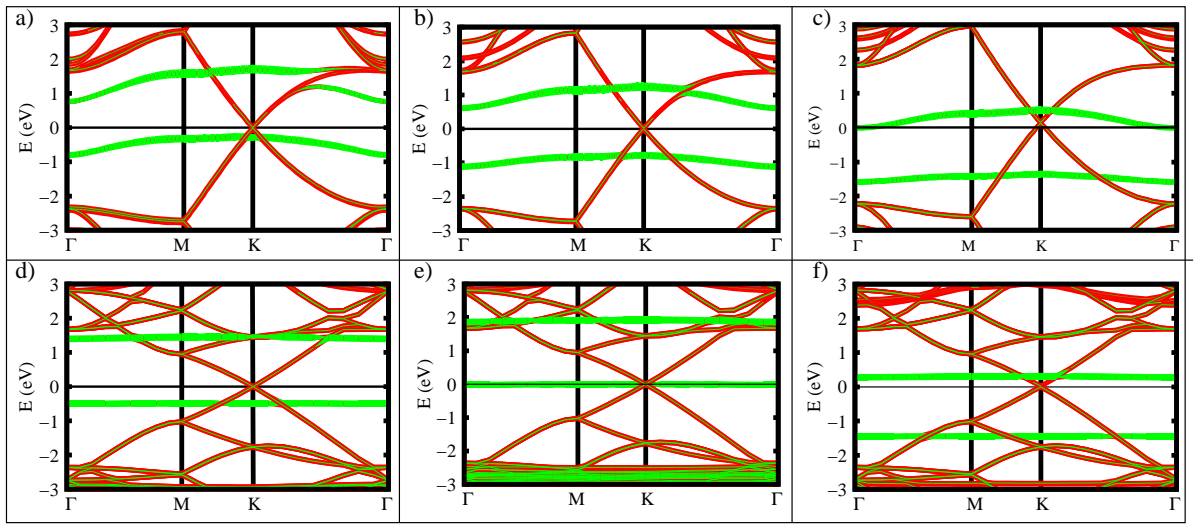}

\caption{\label{fig:substrate_Q0_3_h20_bands}(Color online) Band structures for graphene on defective SiO$_2$ substrates. (a-c) $2\times 2$ and (d-f) $4\times 4$ graphene supercells with every second, (a-c), or eighth, (d-f) surface Si atom forming a $Q_3^0$ defect. The corresponding geometries are shown in Fig. \ref{fig:geom} d-i), respectively. Spin up and down bands are shown at the same time. Contributions at the defect site are marked as green fatbands. a) and d) without water adsorbates. b) with water on top of graphene. c), e), and f) with water between graphene and the substrate. c), f) H$_2$O dipole moment pointing tilted downwards. e) H$_2$O dipole moment pointing upwards.
 }
\end{figure*}

These defects lead to additional states in the vicinity ($\pm
1$eV) around the Fermi level. (See Fig.
\ref{fig:substrate_Q0_3_h20_bands} a) and d).)
The avoided crossing in Fig. \ref{fig:substrate_Q0_3_h20_bands} a)
indicates significant hybridization of the defect and the graphene
bands and demonstrate the impurity state's importance for
conduction electron scattering in graphene. In the pure SiO$_2$
graphene model systems (Fig. \ref{fig:substrate_Q0_3_h20_bands} a)
and d)), the impurity bands do not cross the Fermi level. This
situation can be changed by water adsorbates, which may sit either
in between graphene and the substrate (Fig.
\ref{fig:substrate_Q0_3_h20_bands} c), e) and f) or on top of
graphene (Fig. \ref{fig:substrate_Q0_3_h20_bands} b)). In all
cases, the electrostatic dipole moment of the water adsorbates
comes along with strong local electrostatic fields, which allow to
shift the impurity bands significantly with respect to the
graphene bands. As Fig. \ref{fig:substrate_Q0_3_h20_bands}
demonstrates, this shift strongly depends on the place of
adsorption and the orientation of the water molecules leading,
either to hole doping (Fig. \ref{fig:substrate_Q0_3_h20_bands} c)
and f)) or electron doping (Fig.
\ref{fig:substrate_Q0_3_h20_bands} e)).

This strong effect of water is very general and not limited
to the examples shown in Fig. \ref{fig:substrate_Q0_3_h20_bands}.
Similar effects of water can be found for graphene on the
(0001)-plane of $\alpha$-quartz or $\beta$-cristobalite with
Q$^1_4$ defects. Although the interaction mechanisms of water,
SiO$_2$ and graphene presented here are not exhaustive, the
comparison of water adsorption on perfect free standing graphene
to water adsorption on graphene lying on a (defective) SiO$_2$
substrate allows the following conclusions: Perfect free standing
graphene may have water adsorbates on top but its electronic
transport properties are insensitive against perturbations by the
water adsorbates. Single molecules will not create any impurity
states close to the Dirac point. For obtaining doping effects, one
needs highly ordered H$_2$O cluster or ice structures.

The substrate changes the situation completely. The dipole moments
of H$_2$O adsorbates cause local electrostatic fields that can
shift the substrate's defect states with respect to the graphene
electrons and cause doping. The hybridization of substrate defect
states with the graphene bands can be reduced by H$_2$O in between
graphene and the substrate leading to less scattering of graphene
electrons at defects in the substrate. On the other hand, impurity
bands can be shifted towards the Fermi level by H$_2$O adsorbates,
leading to increased electron scattering. So, for graphene on a
substrate, H$_2$O much more likely affects the electronic
properties than for free standing graphene. The effect of water strongly depends on properties of the substrate like the
amount and type of defects.

This finding might explain experiments on CNTs \cite{KimDai_CNT}:
CNTs on SiO$_2$ substrates are much more sensitive to water in
their environment, than suspended CNTs or CNTs separated from the
SiO$_2$ substrate by poly(methyl methacrylate) (PMMA) coating. In
experiments similar to those in Refs. \cite{morozov08,andrei08}
these effects can be checked for graphene: The creation of
electron/hole puddles and doping on H$_2$O exposure might be
investigated in a study with a similar setup as in Ref.
\cite{andrei08}. We expect much less impact of H$_2$O on the
transport properties for free hanged graphene. Similarly, a
comparison of graphene on hydrophobic substrates and hydrophilic
substrates like PMMA and SiO$_2$ in Ref. \cite{morozov08} should
yield weak and strong response on H$_2$O, respectively.

These findings can be important for the development of
graphene based gas sensing devices: Some molecules like NO$_2$
lead to acceptor states even without a substrate
\cite{WehlingNL08}, while doping effects due to other molecules
like H$_2$O strongly depend on the substrate. This can open the
possibility of selective graphene gas sensors.

The authors are thankful to Andre Geim and Kostya Novoselov for
inspiring discussions. This work was supported by SFB 668
(Germany) and FOM (The Netherlands). Computation time at HLRN is acknowledged.

\bibliography{water_s}

\begin{thebibliography}{20}
\expandafter\ifx\csname natexlab\endcsname\relax\def\natexlab#1{#1}\fi
\expandafter\ifx\csname bibnamefont\endcsname\relax
  \def\bibnamefont#1{#1}\fi
\expandafter\ifx\csname bibfnamefont\endcsname\relax
  \def\bibfnamefont#1{#1}\fi
\expandafter\ifx\csname citenamefont\endcsname\relax
  \def\citenamefont#1{#1}\fi
\expandafter\ifx\csname url\endcsname\relax
  \def\url#1{\texttt{#1}}\fi
\expandafter\ifx\csname urlprefix\endcsname\relax\def\urlprefix{URL }\fi
\providecommand{\bibinfo}[2]{#2}
\providecommand{\eprint}[2][]{\url{#2}}

\bibitem[{\citenamefont{Novoselov et~al.}(2004)\citenamefont{Novoselov, Geim,
  Morozov, Jiang, Zhang, Dubonos, Grigorieva, and
  Firsov}}]{Novoselov_science2004}
\bibinfo{author}{\bibfnamefont{K.~S.} \bibnamefont{Novoselov}},
  \bibinfo{author}{\bibfnamefont{A.~K.} \bibnamefont{Geim}},
  \bibinfo{author}{\bibfnamefont{S.~V.} \bibnamefont{Morozov}},
  \bibinfo{author}{\bibfnamefont{D.}~\bibnamefont{Jiang}},
  \bibinfo{author}{\bibfnamefont{Y.}~\bibnamefont{Zhang}},
  \bibinfo{author}{\bibfnamefont{S.~V.} \bibnamefont{Dubonos}},
  \bibinfo{author}{\bibfnamefont{I.~V.} \bibnamefont{Grigorieva}},
  \bibnamefont{and} \bibinfo{author}{\bibfnamefont{A.~A.}
  \bibnamefont{Firsov}}, \bibinfo{journal}{Science}
  \textbf{\bibinfo{volume}{306}}, \bibinfo{pages}{666} (\bibinfo{year}{2004}).

\bibitem[{\citenamefont{Novoselov et~al.}(2005)\citenamefont{Novoselov, Jiang,
  Schedin, Booth, Khotkevich, Morozov, and Geim}}]{K.S.Novoselov07262005}
\bibinfo{author}{\bibfnamefont{K.~S.} \bibnamefont{Novoselov}},
  \bibinfo{author}{\bibfnamefont{D.}~\bibnamefont{Jiang}},
  \bibinfo{author}{\bibfnamefont{F.}~\bibnamefont{Schedin}},
  \bibinfo{author}{\bibfnamefont{T.~J.} \bibnamefont{Booth}},
  \bibinfo{author}{\bibfnamefont{V.~V.} \bibnamefont{Khotkevich}},
  \bibinfo{author}{\bibfnamefont{S.~V.} \bibnamefont{Morozov}},
  \bibnamefont{and} \bibinfo{author}{\bibfnamefont{A.~K.} \bibnamefont{Geim}},
  \bibinfo{journal}{PNAS} \textbf{\bibinfo{volume}{102}},
  \bibinfo{pages}{10451} (\bibinfo{year}{2005}).

\bibitem[{\citenamefont{Morozov et~al.}(2008)\citenamefont{Morozov, Novoselov,
  Katsnelson, Schedin, Elias, Jaszczak, and Geim}}]{morozov08}
\bibinfo{author}{\bibfnamefont{S.~V.} \bibnamefont{Morozov}},
  \bibinfo{author}{\bibfnamefont{K.~S.} \bibnamefont{Novoselov}},
  \bibinfo{author}{\bibfnamefont{M.~I.} \bibnamefont{Katsnelson}},
  \bibinfo{author}{\bibfnamefont{F.}~\bibnamefont{Schedin}},
  \bibinfo{author}{\bibfnamefont{D.~C.} \bibnamefont{Elias}},
  \bibinfo{author}{\bibfnamefont{J.~A.} \bibnamefont{Jaszczak}},
  \bibnamefont{and} \bibinfo{author}{\bibfnamefont{A.~K.} \bibnamefont{Geim}},
  \bibinfo{journal}{Phys. Rev. Lett.} \textbf{\bibinfo{volume}{100}},
  \bibinfo{pages}{016602} (\bibinfo{year}{2008}).

\bibitem[{\citenamefont{Schedin et~al.}(2007)\citenamefont{Schedin, Geim,
  Morozov, Hill, Blake, Katsnelson, and Novoselov}}]{schedin-gassensors}
\bibinfo{author}{\bibfnamefont{F.}~\bibnamefont{Schedin}},
  \bibinfo{author}{\bibfnamefont{A.~K.} \bibnamefont{Geim}},
  \bibinfo{author}{\bibfnamefont{S.~V.} \bibnamefont{Morozov}},
  \bibinfo{author}{\bibfnamefont{E.~W.} \bibnamefont{Hill}},
  \bibinfo{author}{\bibfnamefont{P.}~\bibnamefont{Blake}},
  \bibinfo{author}{\bibfnamefont{M.~I.} \bibnamefont{Katsnelson}},
  \bibnamefont{and} \bibinfo{author}{\bibfnamefont{K.~S.}
  \bibnamefont{Novoselov}}, \bibinfo{journal}{Nat. Mater.}
  \textbf{\bibinfo{volume}{6}}, \bibinfo{pages}{652} (\bibinfo{year}{2007}).

\bibitem[{\citenamefont{Wehling et~al.}(2008)\citenamefont{Wehling, Novoselov,
  Morozov, Vdovin, Katsnelson, Geim, and Lichtenstein}}]{WehlingNL08}
\bibinfo{author}{\bibfnamefont{T.}~\bibnamefont{Wehling}},
  \bibinfo{author}{\bibfnamefont{K.}~\bibnamefont{Novoselov}},
  \bibinfo{author}{\bibfnamefont{S.}~\bibnamefont{Morozov}},
  \bibinfo{author}{\bibfnamefont{E.}~\bibnamefont{Vdovin}},
  \bibinfo{author}{\bibfnamefont{M.}~\bibnamefont{Katsnelson}},
  \bibinfo{author}{\bibfnamefont{A.}~\bibnamefont{Geim}}, \bibnamefont{and}
  \bibinfo{author}{\bibfnamefont{A.}~\bibnamefont{Lichtenstein}},
  \bibinfo{journal}{Nano Letters} \textbf{\bibinfo{volume}{8}},
  \bibinfo{pages}{173} (\bibinfo{year}{2008}).

\bibitem[{\citenamefont{Leenaerts et~al.}(2008)\citenamefont{Leenaerts,
  Partoens, and Peeters}}]{leenaerts}
\bibinfo{author}{\bibfnamefont{O.}~\bibnamefont{Leenaerts}},
  \bibinfo{author}{\bibfnamefont{B.}~\bibnamefont{Partoens}}, \bibnamefont{and}
  \bibinfo{author}{\bibfnamefont{F.~M.} \bibnamefont{Peeters}},
  \bibinfo{journal}{Phys. Rev. B} \textbf{\bibinfo{volume}{77}},
  \bibinfo{eid}{125416} (\bibinfo{year}{2008}).

\bibitem[{\citenamefont{Zhao et~al.}(2002)\citenamefont{Zhao, Buldum, Han, and
  Lu}}]{Zhao_CNT_H2O_02}
\bibinfo{author}{\bibfnamefont{J.}~\bibnamefont{Zhao}},
  \bibinfo{author}{\bibfnamefont{A.}~\bibnamefont{Buldum}},
  \bibinfo{author}{\bibfnamefont{J.}~\bibnamefont{Han}}, \bibnamefont{and}
  \bibinfo{author}{\bibfnamefont{J.~P.} \bibnamefont{Lu}},
  \bibinfo{journal}{Nanotechnology} \textbf{\bibinfo{volume}{13}},
  \bibinfo{pages}{195} (\bibinfo{year}{2002}).

\bibitem[{\citenamefont{Moser et~al.}(2008)\citenamefont{Moser, Verdaguer,
  Jimenez, Barreiro, and Bachtold}}]{bachtold}
\bibinfo{author}{\bibfnamefont{J.}~\bibnamefont{Moser}},
  \bibinfo{author}{\bibfnamefont{A.}~\bibnamefont{Verdaguer}},
  \bibinfo{author}{\bibfnamefont{D.}~\bibnamefont{Jimenez}},
  \bibinfo{author}{\bibfnamefont{A.}~\bibnamefont{Barreiro}}, \bibnamefont{and}
  \bibinfo{author}{\bibfnamefont{A.}~\bibnamefont{Bachtold}},
  \bibinfo{journal}{Appl. Phys. Lett.} \textbf{\bibinfo{volume}{92}},
  \bibinfo{pages}{123507} (\bibinfo{year}{2008}).

\bibitem[{\citenamefont{Perdew et~al.}(1992)\citenamefont{Perdew, Chevary,
  Vosko, Jackson, Pederson, Singh, and Fiolhais}}]{Perdew:PW91}
\bibinfo{author}{\bibfnamefont{J.~P.} \bibnamefont{Perdew}},
  \bibinfo{author}{\bibfnamefont{J.~A.} \bibnamefont{Chevary}},
  \bibinfo{author}{\bibfnamefont{S.~H.} \bibnamefont{Vosko}},
  \bibinfo{author}{\bibfnamefont{K.~A.} \bibnamefont{Jackson}},
  \bibinfo{author}{\bibfnamefont{M.~R.} \bibnamefont{Pederson}},
  \bibinfo{author}{\bibfnamefont{D.~J.} \bibnamefont{Singh}}, \bibnamefont{and}
  \bibinfo{author}{\bibfnamefont{C.}~\bibnamefont{Fiolhais}},
  \bibinfo{journal}{Phys. Rev. B} \textbf{\bibinfo{volume}{46}},
  \bibinfo{pages}{6671} (\bibinfo{year}{1992}).

\bibitem[{\citenamefont{Perdew et~al.}(1996)\citenamefont{Perdew, Burke, and
  Ernzerhof}}]{PBE}
\bibinfo{author}{\bibfnamefont{J.~P.} \bibnamefont{Perdew}},
  \bibinfo{author}{\bibfnamefont{K.}~\bibnamefont{Burke}}, \bibnamefont{and}
  \bibinfo{author}{\bibfnamefont{M.}~\bibnamefont{Ernzerhof}},
  \bibinfo{journal}{Phys. Rev. Lett.} \textbf{\bibinfo{volume}{77}},
  \bibinfo{pages}{3865} (\bibinfo{year}{1996}).

\bibitem[{\citenamefont{Kresse and Hafner}(1994)}]{Kresse:PP_VASP}
\bibinfo{author}{\bibfnamefont{G.}~\bibnamefont{Kresse}} \bibnamefont{and}
  \bibinfo{author}{\bibfnamefont{J.}~\bibnamefont{Hafner}},
  \bibinfo{journal}{J. Phys.: Condes. Matter} \textbf{\bibinfo{volume}{6}},
  \bibinfo{pages}{8245} (\bibinfo{year}{1994}).

\bibitem[{\citenamefont{Bl\"ochl}(1994)}]{Bloechl:PAW1994}
\bibinfo{author}{\bibfnamefont{P.~E.} \bibnamefont{Bl\"ochl}},
  \bibinfo{journal}{Phys. Rev. B} \textbf{\bibinfo{volume}{50}},
  \bibinfo{pages}{17953} (\bibinfo{year}{1994}).

\bibitem[{\citenamefont{Kresse and Joubert}(1999)}]{Kresse:PAW_VASP}
\bibinfo{author}{\bibfnamefont{G.}~\bibnamefont{Kresse}} \bibnamefont{and}
  \bibinfo{author}{\bibfnamefont{D.}~\bibnamefont{Joubert}},
  \bibinfo{journal}{Phys. Rev. B} \textbf{\bibinfo{volume}{59}},
  \bibinfo{pages}{1758} (\bibinfo{year}{1999}).

\bibitem[{\citenamefont{Lankau}(2004)}]{Lankau:2004}
\bibinfo{author}{\bibfnamefont{T.}~\bibnamefont{Lankau}},
  \emph{\bibinfo{title}{A computational analysis of hydrogen-bonded networks}},
  \bibinfo{address}{Department of Chemistry, University of Hamburg}
  (\bibinfo{year}{2004}), \bibinfo{note}{habilitation thesis}.

\bibitem[{\citenamefont{Meng et~al.}(2007)\citenamefont{Meng, Kaxiras, and
  Zhang}}]{Kaxiras07}
\bibinfo{author}{\bibfnamefont{S.}~\bibnamefont{Meng}},
  \bibinfo{author}{\bibfnamefont{E.}~\bibnamefont{Kaxiras}}, \bibnamefont{and}
  \bibinfo{author}{\bibfnamefont{Z.}~\bibnamefont{Zhang}}, \bibinfo{journal}{J.
  Chem. Phys.} \textbf{\bibinfo{volume}{127}}, \bibinfo{pages}{244710}
  (\bibinfo{year}{2007}).

\bibitem[{\citenamefont{Carrier et~al.}(2001)\citenamefont{Carrier, Lewis, and
  Dharma-wardana}}]{cristobalite}
\bibinfo{author}{\bibfnamefont{P.}~\bibnamefont{Carrier}},
  \bibinfo{author}{\bibfnamefont{L.~J.} \bibnamefont{Lewis}}, \bibnamefont{and}
  \bibinfo{author}{\bibfnamefont{M.~W.~C.} \bibnamefont{Dharma-wardana}},
  \bibinfo{journal}{Phys. Rev. B} \textbf{\bibinfo{volume}{64}},
  \bibinfo{pages}{195330} (\bibinfo{year}{2001}).

\bibitem[{\citenamefont{Wilson and Walsh}(2000)}]{wilson:9180}
\bibinfo{author}{\bibfnamefont{M.}~\bibnamefont{Wilson}} \bibnamefont{and}
  \bibinfo{author}{\bibfnamefont{T.~R.} \bibnamefont{Walsh}},
  \bibinfo{journal}{J. Chem. Phys.} \textbf{\bibinfo{volume}{113}},
  \bibinfo{pages}{9180} (\bibinfo{year}{2000}).

\bibitem[{\citenamefont{Wright and Leadbetter}({1975})}]{cristobalite_wright}
\bibinfo{author}{\bibfnamefont{A.}~\bibnamefont{Wright}} \bibnamefont{and}
  \bibinfo{author}{\bibfnamefont{A.}~\bibnamefont{Leadbetter}},
  \bibinfo{journal}{{Philos. Mag.}} \textbf{\bibinfo{volume}{{31}}},
  \bibinfo{pages}{{1391}} (\bibinfo{year}{{1975}}).

\bibitem[{\citenamefont{Kim et~al.}(2003)\citenamefont{Kim, Javey, Vermesh,
  Wang, Li, and Dai}}]{KimDai_CNT}
\bibinfo{author}{\bibfnamefont{W.}~\bibnamefont{Kim}},
  \bibinfo{author}{\bibfnamefont{A.}~\bibnamefont{Javey}},
  \bibinfo{author}{\bibfnamefont{O.}~\bibnamefont{Vermesh}},
  \bibinfo{author}{\bibfnamefont{Q.}~\bibnamefont{Wang}},
  \bibinfo{author}{\bibfnamefont{Y.}~\bibnamefont{Li}}, \bibnamefont{and}
  \bibinfo{author}{\bibfnamefont{H.}~\bibnamefont{Dai}}, \bibinfo{journal}{Nano
  Letters} \textbf{\bibinfo{volume}{3}}, \bibinfo{pages}{193}
  (\bibinfo{year}{2003}).

\bibitem[{\citenamefont{Du et~al.}(2008)\citenamefont{Du, Skachko, Barker, and
  Andrei}}]{andrei08}
\bibinfo{author}{\bibfnamefont{X.}~\bibnamefont{Du}},
  \bibinfo{author}{\bibfnamefont{I.}~\bibnamefont{Skachko}},
  \bibinfo{author}{\bibfnamefont{A.}~\bibnamefont{Barker}}, \bibnamefont{and}
  \bibinfo{author}{\bibfnamefont{E.~Y.} \bibnamefont{Andrei}},
  \bibinfo{journal}{Nat. Nano.}  (\bibinfo{year}{2008}), \bibinfo{note}{doi:
  10.1038/nnano.2008.199}.

\end{thebibliography}

\end{document}